\documentclass[epj]{svjour}
\usepackage{graphics}
\usepackage{axodraw}
\usepackage{graphicx, subfigure}
\usepackage{slashed}
\begin{document}
\title{Large Hadron Collider constraints on a light baryon number violating sbottom coupling to a top and a light quark}
\author{B. C. Allanach \and S. A. Renner
}                     

%
\institute{DAMTP, CMS, University of Cambridge, Wilberforce Road, Cambridge, CB3 0WA, United Kingdom}
\date{Received: date / Revised version: date}
%
\abstract{
We investigate a model of $R$-parity violating (RPV) supersymmetry in which
the right-handed sbottom is the lightest supersymmetric particle, and a
baryon number violating coupling involving a top is the only non-negligible
RPV coupling.  
This model evades proton decay and flavour constraints. 
We consider in turn each of the couplings $\lambda''_{313}$ and
$\lambda''_{323}$ as the only non-negligible RPV coupling, and 
we recast a recent LHC measurement (CMS top transverse 
momentum $p_T(t)$
spectrum) and a LHC search
(ATLAS multiple jet resonance search) in the form of 
constraints on
the mass-coupling parameter planes. 
We delineate a large
region in the parameter space of the mass of the sbottom ($m_{\tilde{b}_R}$)
and the 
$\lambda''_{313}$ coupling that is ruled out by the measurements, as well as a
smaller region in the parameter space of $m_{\tilde{b}_R}$ and
$\lambda''_{323}$. 
A certain region of the $m_{\tilde{b}_R} \mathrm{-} \lambda''_{313}$ parameter space was
previously found to successfully explain 
the anomalously large $t \bar t$ forward backward asymmetry measured by
Tevatron experiments.
This entire region 
is now excluded at the 95$\%$ confidence level (CL) by CMS measurements of the $p_T(t)$ spectrum.
We also present $p_T(t\bar{t})$ distributions of the
Tevatron $t \bar t$ forward-backward asymmetry for this model.  
} 
\titlerunning{LHC constraints on a light baryon number violating sbottom}
\maketitle
\section{Introduction}
\label{intro}

Supersymmetry (SUSY) is a beyond the Standard Model (BSM) theory that answers
some 
of the unsolved questions of the Standard Model. In particular, weak scale
SUSY provides a solution to the hierarchy problem, which is the problem of
explaining how the Higgs boson mass is stable under radiative corrections
which would otherwise tend to bring it up to huge values in the absence of any
new physics beyond the Standard Model. 
However, there has been no significant evidence for supersymmetry so far at
the LHC. 
 One possible reason for this might be that most of the LHC searches have been
 looking for $R$-parity conserving supersymmetry, which implies a stable
 lightest supersymmetric particle (LSP). 
This LSP would escape the detector undetected, and so searches for this
variety of supersymmetry at the LHC rely on signatures with large missing
transverse momentum. Stringent cuts on the missing transverse momentum are
usually imposed for these analyses. 
 However, if supersymmetry is instead $R$-parity violating (RPV), it can evade these searches because the LSP
 is not stable and so there is no large missing transverse momentum. 
One argument offered for $R$-parity conservation is that it ensures that the
proton is 
stable, but RPV SUSY can also avoid getting into trouble with lower bounds on
proton lifetimes if either baryon number or lepton number is violated, but not
both (proton decay would rely on both being present). 
Recently, it has also been realised that, by considering flavour symmetries
and adding some extra fields charged under such symmetries, a baryon number
violating model may also be 
consistent with stable dark matter constraints~\cite{Batell:2013zwa}. 
Depending on the flavour structure of the baryon number violating couplings,
the gravitino has 
been shown to be a viable dark matter candidate in the $R$-parity violating
MSSM with lifetimes long enough to evade certain bounds~\cite{Lola:2008bk}. 
Thus another argument for $R$-parity conservation (that it guarantees a dark
matter candidate) is seen to be avoidable. 

 If only baryon number violating operators are present, then decays of
 superpartners will produce jets, which might hide amongst large quantum
 chromodynamics (QCD)
 backgrounds at the LHC. 
The general difficulty of discovering baryon number violating SUSY amongst QCD
backgrounds is a well-known one; many papers have discussed this problem and
suggested methods involving studying jet substructure for distinguishing jets
produced through BSM processes~\cite{Duchovni:2013bca,Butterworth:2009qa}. 
 Other suggested analyses have relied on leptons produced in sparticle
 cascades (for example, Ref.~\cite{Allanach:2001xz}).
The tendency of baryon number violating SUSY to ``hide" in QCD backgrounds,
along with the fact that it is expected that third generation squarks should
be light to make the theory more natural~\cite{Brust:2011tb}, has led to the
suggestion that baryon number violating SUSY with light third generation
squarks should be the next new physics scenario to search for, given the lack
of SUSY signals at the LHC so
far~\cite{Allanach:2012vj,Asano:2012gj,Berger:2013sir,Han:2012cu,Durieux:2013uqa}.   

The RPV superpotential within the minimal supersymmetric standard model (MSSM) contains the $B$-violating term 
\begin{equation}
\mathrm{W}=\frac{1}{2}\lambda''_{ijk}U_i^cD_j^cD_k^c,
\end{equation}
where $i,j,k$ are family indices, $U_i^c$ and $D_j^c$ are chiral superfields
containing the 
charge-conjugated right-handed up and down 
type quarks respectively, and we have suppressed gauge indices. The
couplings $\lambda''_{ijk}$ are antisymmetric in the last two 
indices due to the $SU(3)$ colour structure. This superpotential term can be
rewritten in terms of the component fields as 
\begin{equation}
L_{\not{B}}=\lambda''_{ijk}(u_i^cd_j^c\tilde{d}_k^*+u_i^c\tilde{d}_j^*d_k^c+\tilde{u}_i^*d_j^cd_k^c) + h.c.,
\end{equation}
where lower case fields are left-handed Weyl fermions unless they have a
tilde, in which case they are scalars. 
If third generation squarks are light, couplings of third
generation squarks to quarks in the proton, i.e.\ couplings of the form
$\lambda''_{3jk}$ or $\lambda''_{jk3}$, will provide new physics signals at
the LHC. 

Some recent works have built RPV models with minimal flavour
violation~\cite{Nikolidakis:2007fc,Csaki:2013we,Csaki:2011ge,Krnjaic:2012aj} or
product group unification~\cite{biplob} in order to provide natural models that 
evade LHC constraints more 
easily than $R-$parity conserving ones. General features of these models
include a $U_i^cD_j^cD_k^c$ 
operator involving a top (s)quark as the dominant RPV operator, and a
flavour mass hierarchy 
which predicts one of the third generation squarks as a likely LSP\@. The
set up we investigate has these features.

In this paper we will consider the RPV couplings $\lambda''_{313}$ and
$\lambda''_{323}$, which are involved in the vertices shown in
Figure~\ref{fig:vertices}. We will consider each coupling separately, while
setting the other, and all remaining RPV couplings, to zero. 
 We will assume that the right-handed sbottom is the (unstable) LSP, and work
 in a simplified model in which all other superpartners are set to have very
 large masses. The reason for this assumption is simplicity of the parameter
 space: the only 
 relevant parameters for our model are the sbottom mass $m_{\tilde{b}_R}$ and
 the RPV coupling $\lambda''_{313}$ or $\lambda''_{323}$. Our analysis should
 cover a wide range of cases where 
 various sparticles are brought down in mass, but do not result in significant
 top production. One significant caveat could be the case in which gluinos and
 stops are lighter than 1.5 TeV, since then the production of $\tilde g
 \tilde g$, where each gluino decays via first SUSY QCD $\tilde g \rightarrow
 t \tilde t$ followed by the RPV decay of $\tilde t$ into two jets would
 result in significant additional inclusive top production, and affect our
 results. 
 This however 
 would depend upon the branching ratio of the gluino decay into stops: if this
 were small, then our analysis could still apply. 
 {\rm A priori}, it is
 important that the 
 sbottom is the LSP in our scenario, otherwise competing $R$-parity conserving
 decays of the sbottom could play a role, possibly weakening our
 constraints. However, we shall return to this point later, arguing that, to a
 good approximation, our analysis should be insensitive to the identity of the
 LSP. 
 The RPV couplings we consider will have large $\sim {\mathcal O}(1)$
 magnitudes; therefore our analysis should still hold in the presence of other
 $B$-violating RPV couplings that are small compared to this (i.e.\ $<0.3$ or
 so).
In general, there are flavour constraints on RPV couplings coming from
measurements of flavour-changing neutral currents (FCNCs) and meson
mixing~\cite{Giudice:2011ak}. These imply that other $\lambda''_{ijk}$
couplings must be small, for example the particularly strict bound
$\lambda''_{313} \lambda''_{323} < 0.01$ coming from $K^0-\bar K^0$
mixing constraints for sparticle masses less than 1 TeV~\cite{Slavich:2000xm}. 
However, if we assume that there is only one
non-negligible real RPV coupling, these constraints are evaded because 
no tree-level FCNCs are induced. Electric dipole moment 
constraints~\cite{Giudice:2011ak}, which can become important if there are
several non-negligible complex RPV couplings, are also not constraining here. 
In general there are strong constraints coming from atomic parity
violation 
measurements, for example in Cesium ($^{133}$Cs)~\cite{Gresham:2012wc}. 
But Dupuis and Cline have pointed out in their paper~\cite{Dupuis:2012is} that
these 
constraints can be evaded if there is a sizeable amount of $\tilde t$-squark
mixing, 
because two contributing diagrams will then cancel each other, allowing the
model to pass the constraints coming from atomic parity violation. While we
have set the stops to be heavy for our analysis, they could be made lighter to
satisfy the atomic parity violation constraints while not significantly
affecting our predictions. 

\begin{figure}
\begin{centering}
\subfigure[$\lambda''_{313}$]{
\begin{picture}(50,60)(0,0)
\DashArrowLine(0,35)(30,35){4}
\ArrowLine(50,60)(30,35)
\ArrowLine(50,10)(30,35)
\Text(2,40)[bl]{$\tilde{b}_R$}
\Text(38,55)[bl]{$\bar{t}$}
\Text(33,10)[bl]{$\bar{d}$}
\end{picture}}
\hspace{15mm}
\subfigure[$\lambda''_{323}$]{
\begin{picture}(50,60)(0,0)
\DashArrowLine(0,35)(30,35){4}
\ArrowLine(50,60)(30,35)
\ArrowLine(50,10)(30,35)
\Text(2,40)[bl]{$\tilde{b}_R$}
\Text(38,55)[bl]{$\bar{t}$}
\Text(33,10)[bl]{$\bar{s}$}
\end{picture}}

\caption{Relevant vertices involving the $\lambda''_{313}$ and $\lambda''_{323}$ couplings}
\label{fig:vertices}
\end{centering}
\end{figure}
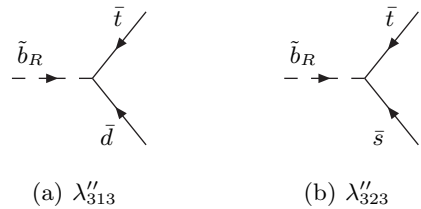

\begin{figure}
\begin{centering}
\subfigure{
\begin{picture}(80,80)(0,0)
\SetColor{Black}
\ArrowLine(0,5)(40,15)
\ArrowLine(80,5)(40,15)
\ArrowLine(40,50)(0,60)
\ArrowLine(40,50)(80,60)
\DashArrowLine(40,50)(40,20){3}
\Text(10,15)[bl]{$d$}
\Text(10,45)[bl]{$\bar{d}$}
\Text(65,15)[bl]{$\bar{t}$}
\Text(65,45)[bl]{$t$}
\Text(50,30)[b]{$\tilde{b}_R$}
\end{picture}{} }
\caption{Diagram of a BSM matrix element producing a $t\bar{t}$ pair (of order
  $\lambda''^2_{313}$).}
\label{fig:LO}
\end{centering}
\end{figure}
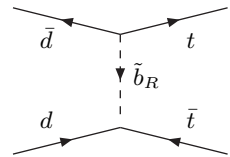
The D$\slashed{0}$ and CDF experiments at the Tevatron have measured a forward-backward
asymmetry in $t\bar{t}$ events~\cite{Dzero:2008,cdf:2008,Aaltonen:2011kc,Abazov:2011rq}. This is not
explained by the 
Standard Model alone, which predicts a significantly smaller value for the
asymmetry \cite{Ahrens:2011uf}. Many proposals were offered for new physics
scenarios that could explain the enhanced asymmetry, see
Refs.~\cite{Kamenik:2011wt,Westhoff:2011tq,AguilarSaavedra:2012ma,Gross:2012bz} for
some examples. 
 In 2012, Allanach and Sridhar proposed one possible explanation for
 this enhanced $t\bar{t}$  asymmetry using RPV supersymmetry
 \cite{Allanach:2012tc}. They showed that an extra diagram contributing to
 $t\bar{t}$ production involving t-channel exchange of a right-handed sbottom
 which couples to top and down quarks via the $\lambda''_{313}$ coupling (as
 shown in Figure~\ref{fig:LO}) could produce an asymmetry which agrees
 with the Tevatron measurements. They checked their model against measurements
 of the $t\bar{t}$ charge asymmetry at the LHC \cite{ATLAS:2012an} and total
 cross-section measurements for a range of values of the sbottom coupling
 $\lambda''_{313}$ (to right-handed down and top quarks) and sbottom mass
 $m_{\tilde{b}_R}$, and found an allowed region for the model in this
 parameter space.  
Around the same time as Allanach and Sridhar's paper, Dupuis and Cline
proposed the same model~\cite{Dupuis:2012is} to explain the $t\bar{t}$
asymmetry, and Hagiwara and Nakamura proposed a very similar model phrased
in terms of diquarks~\cite{Hagiwara:2012gy}. All three papers found
approximately compatible allowed regions in mass-coupling space to explain the
asymmetry. 

In this paper we recast recent LHC measurements in terms of constraints upon
the $m_{\tilde{b}_R}\mathrm{-}\lambda''_{313}$ parameter
space and (separately) the $m_{\tilde{b}_R}\mathrm{-}\lambda''_{323}$ 
parameter space. The disfavoured region in the
$m_{\tilde{b}_R}\mathrm{-}\lambda''_{313}$ parameter space includes Allanach
and Sridhar's region that could explain the $t\bar{t}$ asymmetry whilst
evading other collider constraints.  

The paper is organised as follows: we begin in Section~\ref{sec:ptt} by
looking at the 
$p_T(t\bar{t})$ dependence of the $t\bar{t}$ forward-backward asymmetry as measured by
the CDF experiment at the Tevatron, and compare this to the predictions of the
sbottom model. In Section~\ref{sec:exc} we reinterpret LHC measurements and
calculate excluded regions in mass-coupling parameter spaces of the sbottom. 
We conclude in Section~\ref{sec:conc}. 

\section{Top pair transverse momentum distribution of the forward-backward
  asymmetry \label{sec:ptt}}

Earlier this year the CDF experiment at the Tevatron measured the top quark
forward-backward asymmetry as a function of kinematic properties of the event,
for $t\bar{t}$ events produced by proton-antiproton collisions at a centre of
mass energy of $1.96$ TeV~\cite{Aaltonen:2012it}. 
The $t\bar{t}$ forward-backward asymmetry at CDF, $A_{FB}(t \bar t)$, is defined
\begin{equation}
A_{FB}(t \bar t) = \frac{N(\Delta y > 0)-N(\Delta y < 0)}{N(\Delta y > 0)-N(\Delta y <
  0)}, 
\end{equation}
where $\Delta y = y_t- y_{\bar{t}}$, and $y_t$ and $y_{\bar{t}}$ are the
rapidities of the top and anti-top respectively.

In particular, CDF measured
the forward-backward asymmetry as a function of the transverse momentum ($p_T$)
of the top anti-top pair. A non-zero $p_T$ occurs when the $t \bar t$ system recoils
against an additional jet, for example. This measurement gives new information
to compare to different BSM models which attempt to explain the
forward-backward asymmetry. 
In fact, both colour octet (for example axigluon exchange) models 
and colour singlet models (for example, $Z'$ exchange) were
recently shown to have rather flat differential distributions of $A_{FB}(t
\bar t)$ with $p_T(t \bar t)$~\cite{Webber:2013}. The
predictions from $t$-channel colour anti-triplet exchange have not 
appeared in the literature, and so we provide them here. 

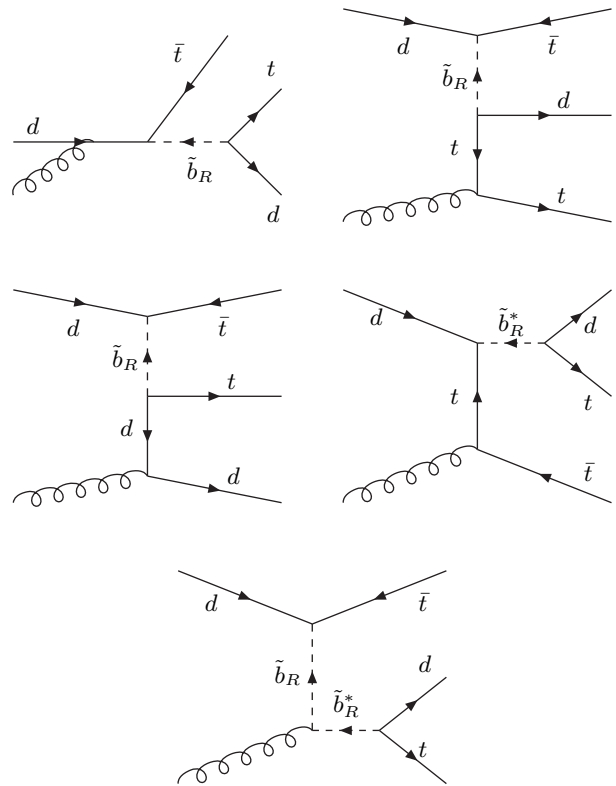
\begin{figure}
\begin{centering}
\subfigure{
\begin{picture}(100,100)(0,0)
\ArrowLine(0,30)(50,30)
\Gluon(0,10)(30,30){3}{4}
\ArrowLine(80,70)(50,30)
\DashArrowLine(80,30)(50,30){3}
\ArrowLine(80,30)(100,10)
\ArrowLine(80,30)(100,50)
\Text(5,33)[bl]{$d$}
\Text(65,15)[bl]{$\tilde{b}_{R}$}
\Text(95,0)[bl]{$d$}
\Text(95,55)[bl]{$t$}
\Text(60,60)[bl]{$\bar{t}$}
\end{picture}{}}
\hspace{5mm}
\subfigure{
\begin{picture}(100,100)(0,0)
\Gluon(0,0)(50,10){3}{5}
\ArrowLine(0,80)(50,70)
\ArrowLine(100,80)(50,70)
\DashArrowLine(50,40)(50,70){3}
\ArrowLine(50,40)(100,40)
\ArrowLine(50,40)(50,10)
\ArrowLine(50,10)(100,0)
\Text(20,63)[bl]{$d$}
\Text(80,63)[br]{$\bar{t}$}
\Text(37,50)[bl]{$\tilde{b}_R$}
\Text(40,25)[bl]{$t$}
\Text(80,43)[bl]{$d$}
\Text(80,7)[bl]{$t$}
\end{picture}{}}
\hspace{5mm}
\subfigure{
\begin{picture}(100,100)(0,0)
\Gluon(0,0)(50,10){3}{5}
\ArrowLine(0,80)(50,70)
\ArrowLine(100,80)(50,70)
\DashArrowLine(50,40)(50,70){3}
\ArrowLine(50,40)(100,40)
\ArrowLine(50,40)(50,10)
\ArrowLine(50,10)(100,0)
\Text(20,63)[bl]{$d$}
\Text(80,63)[br]{$\bar{t}$}
\Text(37,50)[bl]{$\tilde{b}_R$}
\Text(40,25)[bl]{$d$}
\Text(80,43)[bl]{$t$}
\Text(80,7)[bl]{$d$}
\end{picture}{}}
\hspace{5mm}
\subfigure{
\begin{picture}(100,100)(0,0)
\Gluon(0,0)(50,20){3}{5}
\ArrowLine(0,80)(50,60)
\ArrowLine(50,20)(50,60)
\DashArrowLine(75,60)(50,60){3}
\ArrowLine(75,60)(100,40)
\ArrowLine(75,60)(100,80)
\ArrowLine(100,0)(50,20)
\Text(10,67)[bl]{$d$}
\Text(40,37)[bl]{$t$}
\Text(58,63)[bl]{$\tilde{b}_R^*$}
\Text(90,64)[bl]{$d$}
\Text(90,35)[bl]{$t$}
\Text(90,8)[bl]{$\bar{t}$}
\end{picture}{}}
\subfigure{
\begin{picture}(100,100)(0,0)
\Gluon(0,0)(50,20){3}{5}
\ArrowLine(0,80)(50,60)
\DashArrowLine(50,20)(50,60){3}
\DashArrowLine(75,20)(50,20){3}
\ArrowLine(75,20)(100,40)
\ArrowLine(75,20)(100,0)
\ArrowLine(100,80)(50,60)
\Text(10,65)[bl]{$d$}
\Text(36,37)[bl]{$\tilde{b}_R$}
\Text(90,65)[bl]{$\bar{t}$}
\Text(58,25)[bl]{$\tilde{b}_R^*$}
\Text(90,42)[bl]{$d$}
\Text(90,10)[bl]{$t$}
\end{picture}{}}
\caption{Diagrams producing a $t\bar{t}$ pair as well as an extra down quark
  (order $\alpha_s \lambda''^2_{313}$). Not shown, but included in our
  simulations, are diagrams that can be created from these ones by replacing
  quarks with corresponding anti-quarks, and vice versa.} 
\label{fig:extradown}
\end{centering}
\end{figure}

Here, using {\tt MadGraph5\_v1\_5\_11}~\cite{Alwall:2011uj} with
the FeynRules
\cite{Christensen:2008py} implementation of the RPV MSSM
\cite{Fuks:2012im,Duhr:2011se}, we calculate the
distribution for the RPV SUSY model with $\lambda''_{313}$ as the non-zero RPV
coupling. 
We simulate all processes that produce a $t\bar{t}$ pair plus a jet;
i.e.\ the diagram in Figure~\ref{fig:LO} with emission of an additional
gluon, and also the diagrams in Figure~\ref{fig:extradown}, as well as the
leading order QCD processes for $t \bar t$ plus jet production (and interference
between BSM and QCD diagrams). Our simulations were performed at parton level and we
did not decay tops, nor did we include parton showering. We simulated 25 million
events for each histogram. CDF give their results unfolded to the parton
level, so that they can be directly compared to theoretical parton level
predictions.  

\begin{figure*}
\centering\resizebox{12cm}{!}{
\includegraphics[angle=270]{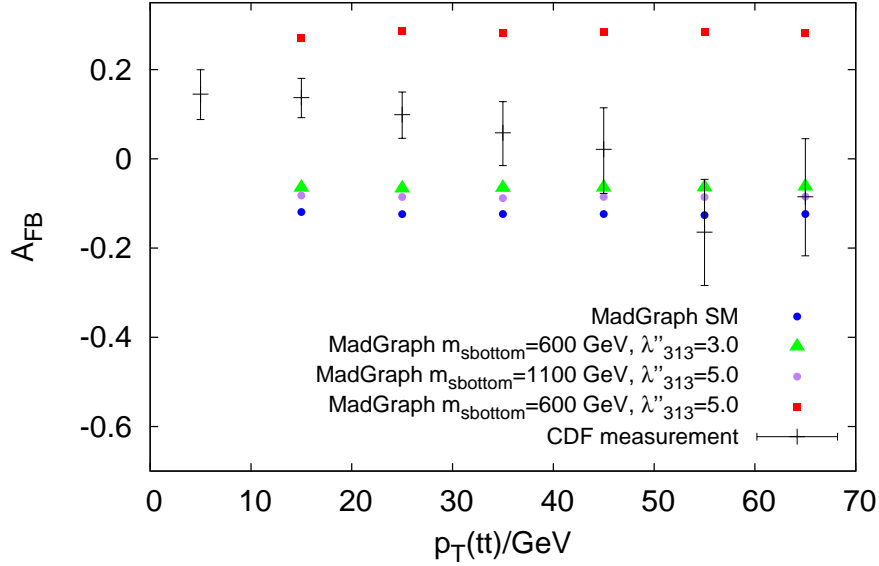}}\\
\caption{Distribution of the top quark forward-backward asymmetry against the
$p_T$ of the $t\bar{t}$ pair at the 1.96 TeV Tevatron. The CDF result is
shown~\cite{Aaltonen:2012it}, 
as well as four distributions calculated by {\tt MadGraph5\_v1\_5\_11}: the
leading order Standard Model $t\bar{t}j$ prediction, and predictions for
$t\bar{t}j$ production via both SM and SUSY processes with a sbottom of mass
600 GeV and $\lambda''_{313}$ coupling of 3.0 or 5.0, and with a sbottom of mass 1100 GeV and $\lambda''_{313}$ coupling of 5.0.}
\label{fig:asymmetry}
\end{figure*}

Figure~\ref{fig:asymmetry} shows the CDF
measurement along with the leading order predictions of {\tt
MadGraph5\_v1\_5\_11}~\cite{Alwall:2011uj} for a sbottom mass of 600 GeV and two
different values of the coupling $\lambda''_{313}$, and for a sbottom mass of 1100 GeV and $\lambda''_{313}=5.0$. These simulations include
leading order 
QCD $t\bar{t}j$ production as well as tree-level processes
involving the RPV sbottom. Our leading order {\tt
MadGraph5\_v1\_5\_11} Standard Model prediction is also shown in
figure~\ref{fig:asymmetry} and is compatible with the recent determination in
Ref.~\cite{Webber:2013}, which uses an independent event generator ({\tt
  HERWIG++}~\cite{Bahr:2008pv}).  

Since the CDF results are unfolded, to compare to
these we did not need to apply any cuts on the simulated $t\bar{t}$ plus jet
system, or on any decay products of the tops. However, {\tt MadGraph5\_v1\_5\_11} can only
simulate tree-level processes - it does
not include loops - so to avoid difficulties with soft jet divergences, we imposed a
lower cut of 10 GeV on the transverse momentum of the jet in our simulated
events. This is why our histograms of {\tt MadGraph5\_v1\_5\_11} predictions in Figure~\ref{fig:asymmetry} do
not include the first bin. 
At a sbottom mass of 600 GeV, the smaller coupling value shown
($\lambda''_{313}=3.0$) falls within the 
region Allanach and Sridhar found which gives the correct value for the total
forward-backward asymmetry, and passed other constraints that were relevant at
the time. A sbottom mass of 1100 GeV and coupling of $5.0$ also falls within this region. The Figure shows that the leading order Standard Model $p_T(t \bar t)$
distribution is fairly flat, in apparent contradiction with the data (the
$\chi^2$-value is $60$ and there are 6 degrees of freedom). 
All of the points in $m_{\tilde{b}_R}-\lambda''_{313}$ parameter space listed
produce a flat distribution, which does not
appear to be mirrored well in the data, 
which has the trend of decreasing $A_{FB}(t \bar t)$ with
increasing top quark pair $p_T$. The prediction of $\lambda''_{313}=3.0$, $m_{\tilde{b}_R}=600$ GeV has a $\chi^2$-value of 35, and that of $\lambda''_{313}=5.0$, $m_{\tilde{b}_R}=1100$ GeV has a $\chi^2$-value of 42. 
The prediction of $\lambda''_{313}=5.0$, $m_{\tilde{b}_R}=600$ GeV is far
above the SM prediction but has a $\chi^2$-value of $61$. 
For 6 degrees of freedom, each of these $m_{{\tilde b}_R}-\lambda''_{313}$
points has a $p$-value of less than $10^{-5}$, as does our Standard Model
result.
We have not included theoretical errors on the {\tt
  MadGraph5\_v1\_5\_11} calculations; of course the $p$-values will alter
somewhat if these are taken into account. 
Throughout this paper, we assume that the likelihood is Gaussian distributed
in the observables and we use two tailed $p$-values to set limits.
 
Colour anti-triplet exchange
thus has a similar status to axigluon or $Z'$ explanations of the $A_{FB}(t
\bar t)$ measurements: $A_{FB}(t \bar t)$ is prediction to be approximately
flat in $p_T$, as is the Standard Model itself.

\begin{figure*}
\subfigure[$m_{\tilde{b}_R} = 250$ GeV]{
\includegraphics[angle=270, width=9cm]{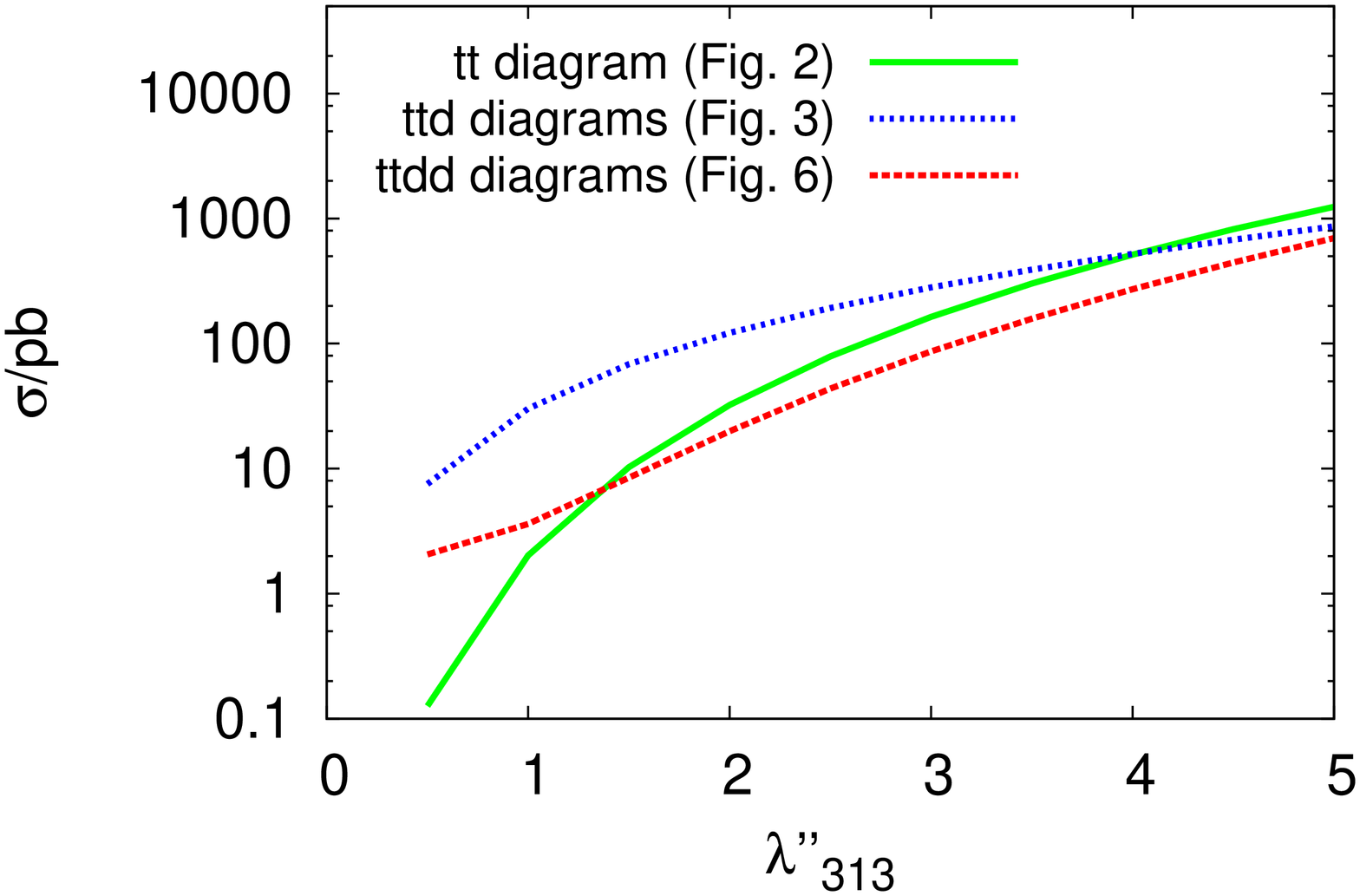}\\
}
\subfigure[$m_{\tilde{b}_R} = 1000$ GeV]{
\includegraphics[angle=270, width=9cm]{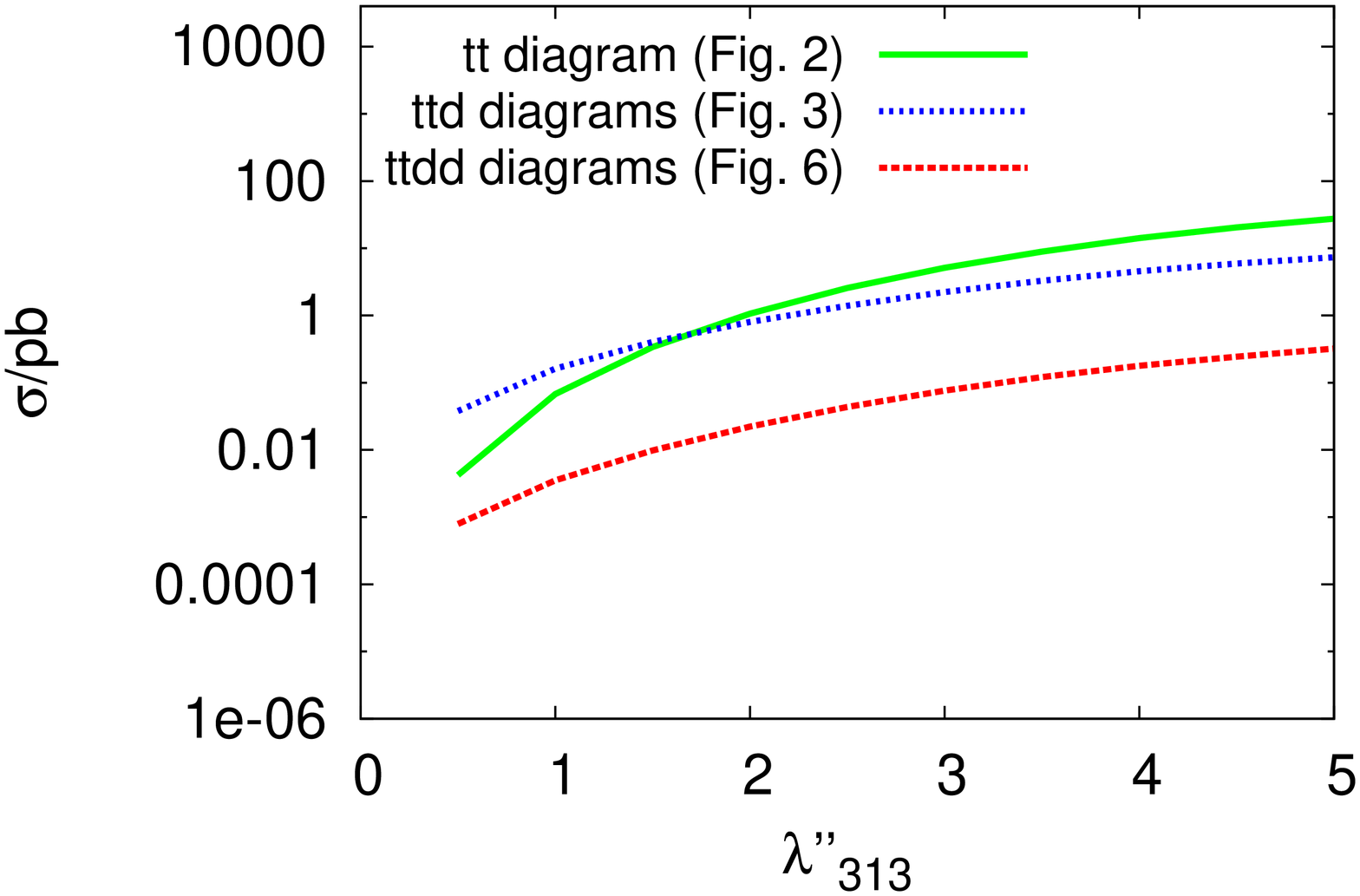}\\
}
\caption{Pure BSM cross-sections as a function of coupling $\lambda''_{313}$ for
  $t\bar{t}$ production events via a sbottom at the 7 TeV LHC: comparing the
  leading-order 
  $t\bar{t}$ process (Figure~\ref{fig:LO}) with $t\bar{t}d$ (Figure~\ref{fig:extradown}) and
  $t\bar{t}d\bar{d}$ (Figure~\ref{fig:extraddbar}) processes, for two values of the sbottom
  mass.} 
\label{fig:xsecs}
\end{figure*}

\section{Recasting an LHC search and an LHC measurement\label{sec:exc}} 

We now sketch the procedure whereby we calculate exclusion regions upon the
relevant parameter space by re-casting LHC measurements and searches in terms
of the RPV light sbottom model.
For each search or measurement to be reinterpreted, experimental observables for
the RPV SUSY 
model were calculated using the matrix element event generator {\tt
 MadGraph5\_v1\_5\_11}~\cite{Alwall:2011uj} assuming a top mass of $m_t$=
172.5 GeV, the 
CTEQ6L1 parton distribution functions (PDFs)~\cite{Pumplin:2002vw} and using
the FeynRules 
\cite{Christensen:2008py} implementation of the RPV MSSM
\cite{Fuks:2012im,Duhr:2011se}. 
 We define $11 \times 11$ grids in $m_{\tilde{b}_R}\mathrm{-}\lambda''_{313}$
and $m_{\tilde{b}_R}\mathrm{-}\lambda''_{323}$ parameter space, simulating 10000
events at each grid point. At different grid points, the only quantities that
are changed in the simulations are the mass, coupling and width of the
sbottom. Predicted observables were interpolated between the grid points. 

Figures~\ref{fig:LO},~\ref{fig:extradown} and~\ref{fig:extraddbar} show the BSM
processes used in our simulations. (We have omitted here, but included in our
simulations, diagrams which can be created from those shown by replacing all
particles with their anti-particles, and vice versa). These are the diagrams for
the case with a non-zero $\lambda''_{313}$ coupling --- for the case with 
a non-zero $\lambda''_{323}$ coupling, every down quark in the diagrams must
be
instead
replaced with a strange quark (and anti-down quarks with anti-strange
quarks). Every process simulated involving the RPV sbottom has a $t\bar{t}$
pair in the final state. 
We found that the cross-sections of the 
$t\bar{t}d$ (Figure~\ref{fig:extradown}) and $t\bar{t}d\bar{d}$ processes
(Figure~\ref{fig:extraddbar}) can be sizeable enough relative to the leading order $t\bar{t}$ diagram (Figure~\ref{fig:LO}),
 in certain regions of
$m_{\tilde{b}_R}-\lambda''_{313}$ parameter space.
They therefore need to be included in our simulations. 
The measurement
of $t \bar t$ production that we use is inclusive, and so these
processes contribute to it. 
Figure~\ref{fig:xsecs} shows the
cross-section for each set of diagrams as a function of the coupling
$\lambda''_{313}$ for two values of the mass of the sbottom. 
For the purposes of illustration, we have not included the pure QCD, nor the
BSM-QCD interference contribution in Figure~\ref{fig:xsecs}, although we include
them as appropriate later when analysing LHC data.
The $t\bar{t}d$ process is seen to have a larger cross-section than $t\bar{t}$
for a sbottom mass of 250 GeV and a $\lambda''_{313}$ coupling less than about 4.0,
and for a sbottom mass of 1100 GeV and coupling less than about 1.5.  The $t\bar{t}d$
process has a gluon replacing either a down or an anti-down quark in the
initial state as compared to the $t\bar{t}$ diagram. The enhanced PDF for
gluons as opposed to anti-downs in the proton at 7-8 TeV, can outweigh the
naive $\alpha_s$ 
suppression of the $t\bar{t}d$ process compared to $t\bar{t}$.
(At the Tevatron, by contrast, valence anti-downs are present in the
collisions, so the $t\bar{t}d$ 
process is a less significant correction to inclusive top pair production than
at the LHC\@. Indeed it was not included in Allanach and Sridhar's simulations.) 

As seen in figure~\ref{fig:xsecs}, the $t\bar{t}d\bar{d}$ BSM process 
is sub-dominant to one of the other two, but it can be 
of the same order as the dominant process, and so we include it in our
simulations. It can have two gluons in its initial state
and so, similarly to the $t\bar{t}d$ process, PDF enhancements can counteract
the naive suppression that is expected for higher order diagrams. 


The cross-sections of processes involving the coupling $\lambda''_{323}$ are
always smaller than equivalent processes involving the coupling
$\lambda''_{313}$, because they require strange quarks and/or anti-quarks in the
initial state, as opposed to downs and/or anti-downs. 
Since there are no valence
strange quarks in protons, but there are valence downs, the strange PDF is
smaller than the down PDF for all values of $x$ (the fraction of the proton 
momentum carried by the interacting parton).
Consequently the excluded regions we found are smaller in the
$m_{\tilde{b}_R}\mathrm{-}\lambda''_{323}$ parameter space than in the
$m_{\tilde{b}_R}\mathrm{-}\lambda''_{313}$ parameter space.

As well as the measurements described below, we also looked at two more LHC searches. One of these was a search for
contact interactions published by the CMS
collaboration~\cite{Chatrchyan:2013muj}. They displayed the inclusive jet $p_T$
spectrum for jets produced in pp collisions at a centre of mass energy of 7
TeV, for jets with a $p_T$ between 507 and 2116 GeV. We tried to produce an
exclusion region for the sbottom in mass-coupling parameter space using this
measurement, but the cross-sections for the
simulated RPV SUSY events were too low (by a factor of about 15) to exclude any
points within either mass-coupling parameter space grid.

We also looked at a recent search by CMS for pair-production of resonances, each 
decaying to a top and a jet~\cite{CMS:asa}. This is obviously relevant to our sbottom, which
decays to a top and either a down or a strange quark. CMS looked for a bump in
the invariant mass of the top and jet in events with two tops and two jets.
They were searching specifically for an excited top which decays into a top and a gluon.
These excited tops are very narrow, whereas our sbottom LSP is much wider,
with a width given by
\begin{equation}
\Gamma = \frac{\lambda''^2_{3i3}(m_{\tilde{b}_R}^2-m_{t}^2)^2}{8 \pi
  m_{\tilde{b}_R}^3}, \label{sbotwidth}
\end{equation}
where $m_t$ is the top quark mass. 
In the CMS paper, a plot is presented that gives the excluded values of the
cross-section of pair production of the resonance as a function of the mass of
the resonance. However, this limit is based on a calculation with 
resonance significantly narrower than ours, and narrower resonances are easier
to see against 
a smoothly decaying background than wider ones, so we cannot justify directly
applying the limits to our model. We cannot create our own exclusion from
their plot of the differential cross-section as a function of the invariant
mass of the top quark plus jet, because then we would need to accurately model
the experimental resolution. 
We have however found one LHC measurement and one LHC search which yield
strong constraints on our model. Below, we describe the measurement first.

\subsection{Differential top quark transverse momentum}

The first excluded regions were calculated using the differential top
transverse momentum distribution in dileptonic $t\bar{t}$ production events as
measured by the CMS collaboration at the LHC~\cite{Chatrchyan:2012saa}. CMS
measured the differential cross-section of $t\bar{t}$ events as a function of
the transverse momentum of the top quarks (including both top and anti-top
quarks) in 
5 $\mathrm{fb}^{-1}$ of proton-proton collisions at a centre of mass energy of $7$ TeV.

\begin{figure}
\begin{centering}
\subfigure{
\begin{picture}(100,100)(0,0)
\SetColor{Black}
\ArrowLine(0,10)(35,25)
\DashArrowLine(70,20)(35,25){4}
\ArrowLine(35,55)(0,70)
\DashArrowLine(35,55)(70,60){4}
\ArrowLine(35,55)(35,25)
\ArrowLine(100,75)(70,60)
\ArrowLine(100,48)(70,60)
\ArrowLine(70,20)(100,5)
\ArrowLine(70,20)(100,32)
\Text(5,18)[bl]{$d$}
\Text(5,55)[bl]{$\bar{d}$}
\Text(28,35)[bl]{$t$}
\Text(52,45)[bl]{$\tilde{b}_R$}
\Text(52,25)[bl]{$\tilde{b}_R^*$}
\Text(85,75)[bl]{$\bar{t}$}
\Text(85,41)[bl]{$\bar{d}$}
\Text(85,0)[bl]{$d$}
\Text(85,30)[bl]{$t$}
\end{picture}{} }
\hspace{5mm}
\subfigure{
\begin{picture}(100,100)(0,0)
\ArrowLine(0,0)(50,0)
\ArrowLine(50,80)(0,80)
\ArrowLine(50,80)(100,80)
\DashArrowLine(50,80)(50,53){3}
\ArrowLine(100,53)(50,53)
\ArrowLine(50,26)(50,53)
\ArrowLine(50,26)(100,26)
\DashArrowLine(50,26)(50,0){3}
\ArrowLine(100,0)(50,0)
\Text(10,66)[bl]{$\bar{d}$}
\Text(10,5)[bl]{$d$}
\Text(37,60)[bl]{$\tilde{b}_R$}
\Text(40,40)[bl]{$t$}
\Text(37,5)[bl]{$\tilde{b}_R$}
\Text(90,73)[bl]{$t$}
\Text(90,55)[bl]{$\bar{d}$}
\Text(90,27)[bl]{$d$}
\Text(90,3)[bl]{$\bar{t}$}
\end{picture}{} }
\hspace{5mm}
\subfigure{
\begin{picture}(120,100)(0,0)
\ArrowLine(0,0)(30,10)
\ArrowLine(30,50)(0,80)
\DashArrowLine(30,50)(30,10){3}
\ArrowLine(30,50)(60,50)
\ArrowLine(100,30)(60,50)
\ArrowLine(80,0)(30,10)
\DashArrowLine(80,70)(60,50){3}
\ArrowLine(80,70)(120,50)
\ArrowLine(80,70)(120,80)
\Text(3,60)[bl]{$\bar{d}$}
\Text(3,8)[bl]{$d$}
\Text(16,22)[bl]{$\tilde{b}_R$}
\Text(40,53)[bl]{$t$}
\Text(55,60)[bl]{$\tilde{b}_R^*$}
\Text(90,23)[bl]{$\bar{d}$}
\Text(70,5)[bl]{$\bar{t}$}
\Text(110,70)[bl]{$t$}
\Text(107,45)[bl]{$d$}
\end{picture}{}}
\hspace{5mm}
\subfigure{
\begin{picture}(100,70)(0,0)
\ArrowLine(0,10)(20,35)
\ArrowLine(20,35)(0,60)
\Gluon(20,35)(45,35){3}{3}
\DashArrowLine(45,35)(70,50){4}
\DashArrowLine(70,20)(45,35){4}
\ArrowLine(100,70)(70,50)
\ArrowLine(100,40)(70,50)
\ArrowLine(70,20)(100,0)
\ArrowLine(70,20)(100,30)
\Text(5,0)[bl]{$q$}
\Text(5,63)[bl]{$\bar{q}$}
\Text(50,50)[bl]{$\tilde{b}_R$}
\Text(50,13)[bl]{$\tilde{b}_R^*$}
\Text(95,75)[bl]{$\bar{t}$}
\Text(95,44)[bl]{$\bar{d}$}
\Text(95,5)[bl]{$t$}
\Text(95,20)[bl]{$d$}
\end{picture}}
\hspace{5mm}
\subfigure{
\begin{picture}(100,100)(0,0)
\Gluon(0,10)(20,35){3}{3}
\Gluon(0,60)(20,35){3}{3}
\Gluon(20,35)(45,35){3}{3}
\DashArrowLine(45,35)(70,50){4}
\DashArrowLine(70,20)(45,35){4}
\ArrowLine(100,70)(70,50)
\ArrowLine(100,40)(70,50)
\ArrowLine(70,20)(100,0)
\ArrowLine(70,20)(100,30)
\Text(50,50)[bl]{$\tilde{b}_R$}
\Text(50,13)[bl]{$\tilde{b}_R^*$}
\Text(95,75)[bl]{$\bar{t}$}
\Text(95,44)[bl]{$\bar{d}$}
\Text(95,5)[bl]{$t$}
\Text(95,20)[bl]{$d$}
\end{picture}}

\caption{Diagrams producing a $t\bar{t}$ pair as well as a down and an
  anti-down (order $\lambda''^4_{313}$ or order $\lambda''^2_{313}\alpha_s^2$). Here, only two order $\lambda''^2_{313}\alpha_s^2$ diagrams are shown, but there are many more diagrams of the same order (all are included in simulations). Also not shown, but included in our
  simulations, are diagrams that can be created from these ones by replacing
  quarks with corresponding anti-quarks, and vice versa.} 
\label{fig:extraddbar}
\end{centering}
\end{figure}
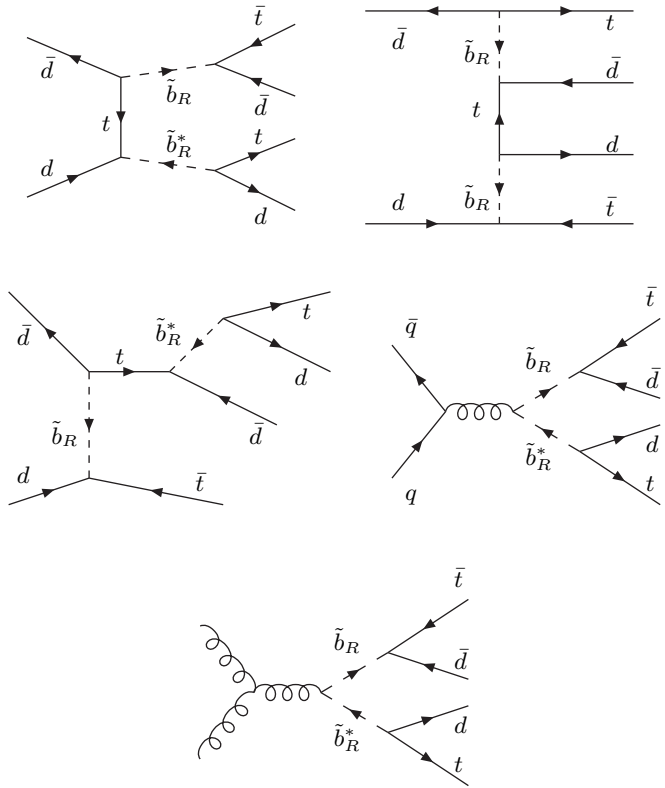

The SUSY $t\bar{t}$ processes that we simulated for the $\lambda''_{313}$
coupling case are shown in Figs~\ref{fig:LO},~\ref{fig:extradown}
and~\ref{fig:extraddbar}. Equivalent diagrams 
with down quarks replaced by strange quarks were simulated for the case
involving the $\lambda''_{323}$ coupling. The leading order Standard Model
$t\bar{t}$ production diagram was also included.

Our simulations were performed at the parton level and we did not decay tops
nor did we include parton showering. The CMS measurement is presented in their
paper 
after having been unfolded to the full $t\bar{t}$ phase space so we are
justified in comparing it directly to parton level $t\bar{t}$ simulated
events without cuts.

\begin{figure}
\resizebox{9cm}{!}{
\includegraphics[angle=270]{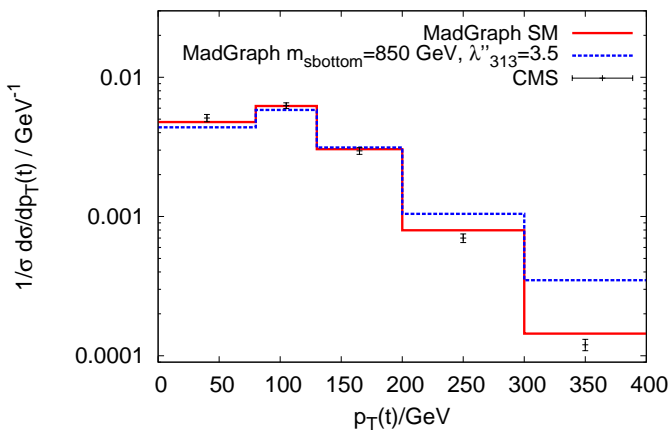}}\\
\caption{Differential distribution of the $p_T$ of the top quarks in
 $t\bar{t}$ 7 TeV LHC events (in which both top and anti-top decay
 leptonically, and  the $p_T$s of both are included in the distribution). The
 CMS measurement and its 
 error bars are shown~\cite{Chatrchyan:2012saa}, as well as two distributions
 calculated by {\tt MadGraph5\_v1\_5\_11}. The {\tt MadGraph5\_v1\_5\_11}
 Standard Model prediction is shown as the solid
 histogram, and the {\tt MadGraph5\_v1\_5\_11} prediction for $t\bar{t}$
 production via Standard Model plus sbottom induced processes as the dotted
 histogram, for sbottoms of mass 850 GeV and $\lambda''_{313}=3.5$.}  
\label{fig:toppT}
\end{figure}

A statistical comparison between measurement and simulation was made using the
CL$_s$ test~\cite{Beringer:1900zz,Read:2002hq}. 
At each point on the parameter space grid, the differential $p_T$ distribution
of the tops was calculated and binned in the same way as in the CMS paper. The
differential distribution of the top $p_T$ is illustrated for the CMS
measurement and 
our 
{\tt MadGraph5\_v1\_5\_11} calculations in Figure~\ref{fig:toppT}.
We see from the figure that the new physics contribution enhances the high
$p_T(t)$ tail.

The ``background-only hypothesis" for the CL$_s$ test was taken to be the NNLO SM prediction, taken from the CMS paper. Then the SM plus sbottom prediction was taken to be the NNLO predicted histogram, minus the MadGraph SM histogram, as shown in Figure~\ref{fig:toppT}, added on to the MadGraph SM plus sbottom histogram (taking into account the differing cross-sections in this sum).
The CL$_s$ test was used, for 4 degrees of freedom, 
to determine which points 
on the grid fell inside the 95\% confidence level exclusion regions, ie. where the value of CL$_s$ is less than 5\% (we have
normalised the area of each histogram in Figure~\ref{fig:toppT} to 1, losing one degree of freedom with
respect to the number of bins).

\begin{figure*}
\centering\resizebox{15cm}{!}{
\includegraphics[angle=270]{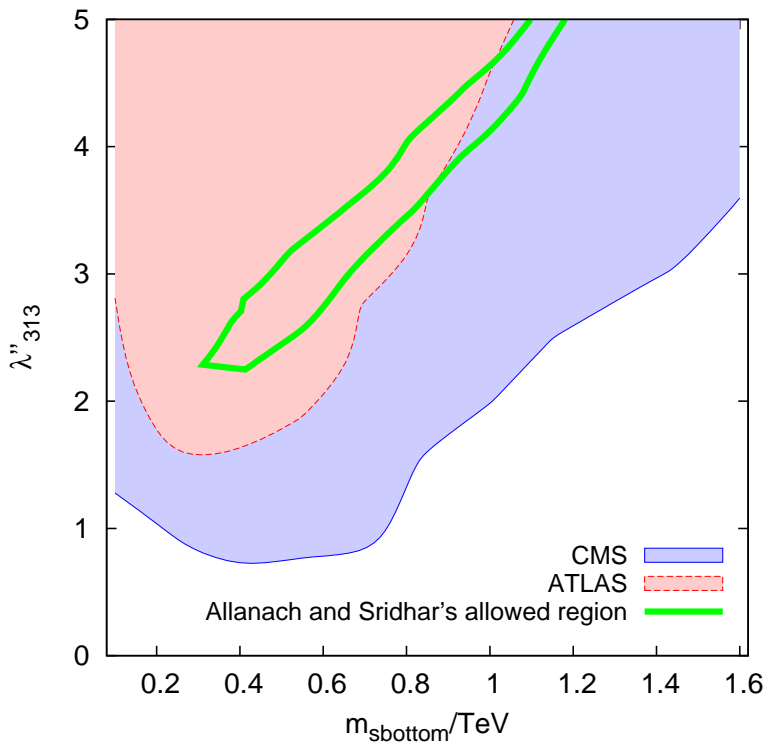}}\\
\caption{$m_{\tilde{b}_R}\mathrm{-}\lambda''_{313}$ parameter space, showing the
95\% CL exclusion regions from (a) the top $p_T$ spectrum measured by CMS
in 5 $\mathrm{fb}^{-1}$ of 7 TeV LHC collisions, marked ``CMS"~\cite{Chatrchyan:2012saa},
and (b) a multiple jet resonance search by ATLAS in 20.3 $\mathrm{fb}^{-1}$ of 8 TeV LHC
collisions~\cite{ATLAS:2}, marked ``ATLAS". }
\label{fig:total313}   
\end{figure*}

\begin{figure*}
\centering\resizebox{15cm}{!}{
\includegraphics[angle=270]{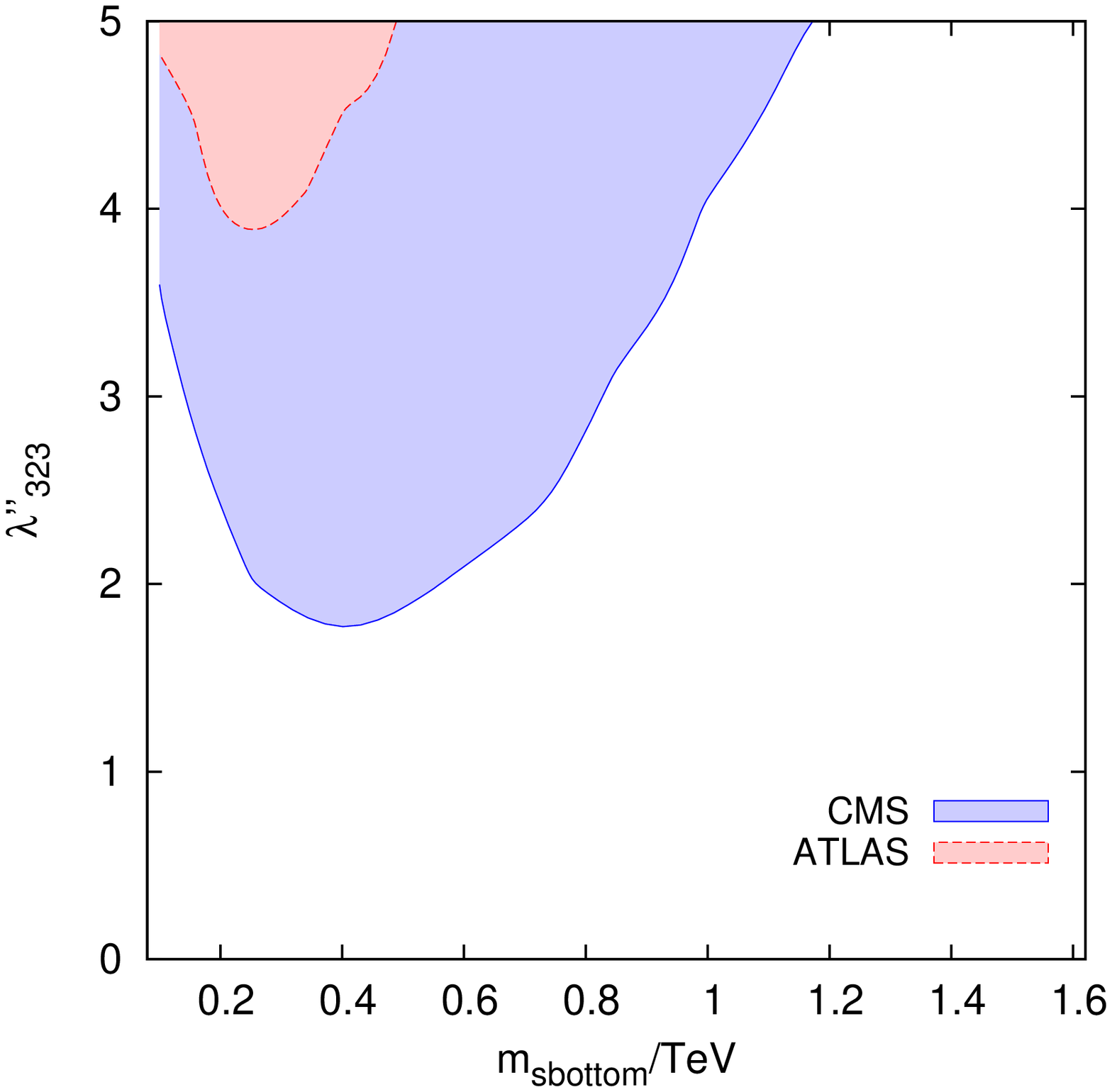}}\\
\caption{$m_{\tilde{b}_R}\mathrm{-}\lambda''_{323}$ parameter space, showing the
95\% CL exclusion  regions from (a) the top $p_T$ spectrum measured by CMS
in 5 $\mathrm{fb}^{-1}$ of 7 TeV LHC collisions, marked ``CMS"~\cite{Chatrchyan:2012saa},
and (b) a multiple jet resonance search by ATLAS in 20.3 $\mathrm{fb}^{-1}$ of 8 TeV LHC
collisions~\cite{ATLAS:2}, marked ``ATLAS". }
\label{fig:total323}
\end{figure*}

Figure~\ref{fig:total313} shows the constraint on the
$m_{\tilde{b}_R}\mathrm{-}\lambda''_{313}$ parameter space coming from the
measurement of the
differential distribution of the top $p_T$ in the dilepton channel, as labelled
in the legend. The region inside the line labelled `Allanach and Sridhar's allowed
region' is taken from Ref.~\cite{Allanach:2012tc}, and is consistent with the
95$\%$ CL
regions of: the CDF and D$\slashed{0}$
data on $A_{FB}(t \bar t)$ for low and high invariant mass bins~\cite{CDF:1},
the total $t \bar t$ production
cross-section~\cite{Aaltonen:2009},
 the CDF differential cross-section with respect to the $t \bar t$ invariant
mass~\cite{Aaltonen:2009iz},
and the ATLAS~\cite{ATLAS:1} and CMS~\cite{CMS:1} $t \bar t$ cross-sections
measured at 7 TeV. It is also consistent with early measurements of the charge
asymmetry at 7 TeV by ATLAS~\cite{ATLAS:2012an} and 
CMS~\cite{Chatrchyan:2011hk}. Much of the higher $\lambda''_{313}$ parameter
space is ruled out by the $p_T(t)$ distribution. 
Figure~\ref{fig:total323} shows the constraints given by the top $p_T$
measurement on $m_{\tilde{b}_R}\mathrm{-}\lambda''_{323}$ parameter space, yielding
significant constraints  (albeit weaker ones than on $\lambda''_{313}$).


\subsection{ATLAS search for pair production of 
  massive particles decaying 
  into several quarks} 
The ATLAS collaboration recently undertook a search for the production of
pairs of 
massive particles, each of which decays into multiple quarks, in 20.3
fb$^{-1}$ of proton-proton
collisions at $\sqrt{s}=8$ TeV at the LHC~\cite{ATLAS:2}. They were looking
in particular for baryon number violating gluinos which decay to 3 or 5 quarks
each. The search 
involved counting the number of events which contained at least 7 jets all with
$p_T>80$ GeV and $|\eta|<2.8$ ($\eta$ is pseudorapidity), and with either 0, 1
or 2 $b$-tags (where the $b$-jets must have $|\eta|<2.5$). Since our signal
contains a $t\bar{t}$ pair in the final state, we 
used the ATLAS 2 $b$-tag event count to calculate an exclusion region. 

Using {\tt MadGraph5\_v1\_5\_11}, we simulated all of the processes shown in
Figures~\ref{fig:extradown} and~\ref{fig:extraddbar} (we excluded the
diagram in Figure~\ref{fig:LO} since it cannot produce 7 partons in the
final state), decaying the tops hadronically. Then for each value of the
sbottom mass 
and coupling values investigated, the cross-section was taken to
be the  
fraction of events that passed the cuts (i.e.\ those which contain at least 7
final-state partons each with $p_T>80$ GeV and $|\eta|<2.8$ of which two are b
quarks with $|\eta|<2.5$) in the simulated event 
samples times the production cross-section, plus
the background estimation given in the ATLAS paper. The number of events
predicted is then this cross-section multiplied by the integrated luminosity.
Using the $\chi^2$ test between the number of
events found in this way and ATLAS's measured number, for each point in
mass-coupling parameter space, we were able to find the regions of $m_{\tilde{b}_R}\mathrm{-}\lambda''_{313}$ and $m_{\tilde{b}_R}\mathrm{-}\lambda''_{323}$ parameter
space that are excluded at 95\% by the ATLAS measurement. These regions are
shown 
in Figure~\ref{fig:total313} and Figure~\ref{fig:total323}.

\section{Conclusions \label{sec:conc}}

We have investigated constraints on a light   sbottom which couples to quarks
via the $R$-parity violating coupling $\lambda''_{313}$ or 
$\lambda''_{323}$. Our constraints complement recent work which focuses on
baryon number violating decays of top
squarks whose mother is a gluino~\cite{Allanach:2012vj,Asano:2012gj,Berger:2013sir,Han:2012cu,Durieux:2013uqa} which leads to the experimentally
advantageous like-sign dilepton signature. 
Using recent LHC measurements, we have ruled out a large
region in $m_{\tilde{b}_R}\mathrm{-}\lambda''_{313}$ parameter space. This
region includes the entire previously allowed parameter space
region~\cite{Hagiwara:2012gy,Allanach:2012tc,Dupuis:2012is} which 
explains the anomalously high $t\bar{t}$ forward-backward asymmetry at the
Tevatron~\cite{Dzero:2008,cdf:2008,Aaltonen:2011kc,Abazov:2011rq}. The excluded
region in 
$m_{\tilde{b}_R}-\lambda''_{323}$ parameter space is smaller, because processes
involving the $\lambda''_{323}$ coupling require strange quarks in the initial
state as opposed to down quarks. 
The associated PDF suppression in the cross-sections of processes involving
the  $\lambda''_{323}$ coupling, relative to those involving similar values of
$\lambda''_{313}$, makes it more difficult to constrain 
$m_{\tilde{b}_R}\mathrm{-}\lambda''_{323}$ parameter space. 

Excluded RPV couplings are rather large (higher than about 0.7), and therefore 
we see that our results should be fairly robust with respect to changes to our
initial simplifying assumption that the sbottom is the LSP\@. If the sbottom
were not the LSP, the worry was that competing $R$-parity conserving decays
would weaken our bounds. While this is in principle true, the $R$-parity
conserving decay 
modes will likely be sub-dominant to the RPV decay modes for couplings higher than about
0.7, and so the effect of having a different LSP on our observables is likely
to be small. For the same reason, $R$-parity conserving contributions to the
sbottom width are likely to be small compared to  Eq.~\ref{sbotwidth}.

Our simulations were performed at the parton level. But we can be confident that
our conclusions are reliable without simulating parton showering,
hadronisation and detectors, because the most constraining
measurement is the CMS top $p_T$ distribution, which was unfolded to the
$t\bar{t}$ level. 

We presented the top pair $p_T$ dependence of the Tevatron forward-backward
asymmetry predicted by this sbottom model (with $\lambda''_{313}$ as the
non-zero RPV coupling). We found it to predict a flat distribution, which does
not fit CDF data well~\cite{Aaltonen:2012it}.  

We have investigated both possibilities for a real $\lambda''_{3j3}$ coupling
for which the sbottom couples to a top. But there are of course other
possibilities for the dominant $\lambda''_{ij3}$ coupling which do not involve
tops. For example the sbottom could couple to an up quark and a strange quark.
In this situation, the sbottom would be more difficult to discover at the
LHC,  since the signal would be hiding in the extremely large jet
background.

We had some trouble finding LHC searches which would be sensitive to our
signal. Most of the SUSY searches are not applicable because
they usually put strong lower cuts on the missing transverse momentum (MET).
They also often veto 
leptons in the event to ensure that the MET does not 
come from a W boson decaying leptonically. This is because they are looking for
stable LSPs which would show up as large MET, and they
want to exclude events where the only MET is due to
a neutrino, since such events constitute a new physics background. Since the
only source of MET in our signal is neutrinos from tops 
decaying leptonically, a large part of our signal does not pass cuts on the
$R$-parity conserving 
SUSY searches and we cannot use them to strongly constrain
the model. ATLAS have 
recently performed a search for $B$-violating operators in RPV supersymmetry,
but the search required kinematically accessible gluinos in the model, and
their signal was same-sign dileptons, neither of which are predicted by our
set up~\cite{ATLAS:2013tma}. 
In this case, for a 100$\%$ branching ratio of $\tilde g \rightarrow
 t bs$, the experimental limit $m_{\tilde g}> 900$ GeV
applies~\cite{ATLAS:2013tma}. The recent recasting~\cite{Berger:2013sir} of
3.95 fb$^{-1}$ 
of an 8 TeV CMS $b$-tags and like-sign lepton search yields $m_{\tilde g}> 800$
GeV~\cite{CMS-PAS-SUS-12-017}. 
Recent searches for RPV SUSY that 
look in particular for lepton number violating operators
\cite{Chatrchyan:2013xsw,ATLAS:3} require more leptons in the final state
than our signal produces, so we cannot 
reinterpret these to put bounds on our model. 
However, we expect precision top measurements to 
better exclude this model in the future, because processes involving these
LSP sbottoms 
alter the differential production cross-section of tops.

\section{Acknowledgements} 
This work has been partially supported by STFC\@. We 
thank M. Schmaltz for the initial idea. Thanks to the Cambridge Supersymmetry
Working Group for helpful discussions. 

\bibliographystyle{epjc} \bibliography{sbottom}

\end{document}